# World-First SEM-based Recovery of Crash EDR Data from the EEPROM of a Severely Damaged SRS Module Using CrashScan

Dr Sergei Skorobogatov
*Cambridge Research and Engineering*
Waterbeach, United Kingdom
sergei.skorobogatov@hushmail.com

Dr Peter Vertal
*Vertal Forensic*
Mlynica, Slovakia
info@vertal.sk

***Abstract*—This paper introduces fully working and affordable approach to data recovery from automotive SRS (airbag control) module after permanent failure of its data storage memory chip. All essential information was successfully extracted with 100% success rate from a real SRS module after severe accident which resulted in both its Processor and Flash EEPROM chip electrically damaged with visible holes in their packages. Further investigation revealed that the damage was likely caused by overvoltage and subsequent high-current-flow through silicon die of chips. This not only caused some internal metal wires to melt and evaporate but also caused partial evaporation of silicon substrate and the created cavity was filled with melted gold from a bonding wire. This reveals the nature of the failure with local temperatures exceeding 3000°C. Despite to the fact that the on-chip EEPROM array was only a few hundred micrometers away from the damaged area, it did not lose any single bit of information. This proved the approach to be resilient for successful data recovery from severely damaged devices in the future.**

**With the donor unit containing the recovered data, standard bench-top EDR extraction was performed using Collision Sciences CrashScan, yielding a full EDR report. The forensic significance of the recovered data, the technical novelty of the methodology, and the complete accident reconstruction are presented and discussed.**

**Until now, if because of an accident the memory chip was damaged, there was no feasible or affordable solution for full data extraction or recovery. With the use of innovative and proprietary sample preparation and imaging techniques, it became possible to extract and verify all the data stored inside chips after permanent mechanical, electrical or fire damage. This could help not only automotive but also medical, aviation and aerospace industries.**

## I. Introduction

SRS (Supplemental Restraint System) module plays very important role in modern cars because of high demand for safety and a high speed they usually travel at. This module is the computer brain that controls your vehicle's crash sensors, airbags, seatbelt pretensioners, and other active safety systems. It is also commonly referred to as the airbag control module, occupant restraint controller, or crash data box. It continuously reads data from impact sensors, wheel speed sensors and seat occupancy detectors. In case of any impact, it instantly calculates crash severity and deploys airbags and seatbelt pretensioners during an accident. It also stores critical vehicle metrics during a crash, acting like a flight "black box". To maintain its continuous functionality, SRS performs self-checks every time you start the car to ensure all safety systems work as intended [1].

SRS module has multiple components on its PCB, but the most important ones are Processor and Non-Volatile Memory (NVM) chip. The processor communicates with car computer, reads data from various sensors and stores processed information inside the memory chip. It is also responsible for triggering active safety systems such as airbags, seatbelt pretensioners and deployable bonnet system. The memory chip, usually SPI Flash EEPROM, preserves very important information about any serious accidents which is crucial for investigations. This data helps to determine the cause of the event and find out whether it was a vehicle failure, driver mistake or something else. If your airbags deployed, the module locks up and stores the crash data. Not only acceleration readings from all sensors within the last few seconds but also steering wheel and pedals positions, as well as the exact timing of airbags deployment and activation of other safety systems.

Data extraction forms an important step in many accident investigations [2]. Standard OBD2 scan tools cannot access SRS and read crash data. You must use special tools to access the unit and read out all the necessary crash data. However, if because of an accident, the module is not accessible via OBD2, there are only few options left. First, the SRS module must be physically removed from the car. Some tools can be directly connected to the module and communicate via CAN, LIN or other interface. In case the module itself was damaged, it might be possible to carefully remove the Flash EEPROM memory chip from its PCB and read out the binary data for further analysis. For this, more sophisticated tools are required because of the raw data stored inside the memory chip. If such tools are unavailable, one can consider reverse engineering the firmware inside the Processor chip. Very often these processors are standard microcontrollers with obscured marking to prevent their reverse engineering [3].

Event Data Recorders (EDR) embedded in modern vehicle Airbag Control Modules (ACM) represent one of the most valuable sources of objective technical evidence in forensic traffic accident reconstruction, as demonstrated in a series of prior publications by the author [4–9], EDR data can be reliably retrieved from a wide range of vehicle types using tools such as Collision Sciences CrashScan, providing pre-crash speed profiles, braking behavior, steering inputs, airbag deployment timelines, and crash pulse characteristics. The forensic evidential value of this data has been validated through controlled crash tests, cross-verification with independent reference systems, and real-world case studies. However, all previously documented EDR extraction methodologies share a common prerequisite: the ACM must be electrically functional, capable of establishing diagnostic communication via the CAN bus, K-Line or Flexray interface. This condition can fail in a variety of real-world scenarios — severe crash-induced structural damage, post-crash fire, submersion, or, as in the present case, an external electrical event that destroys the module's communication circuitry while leaving the physical accident scene and the vehicle otherwise intact. When this condition fails, the forensic expert faces a fundamental impasse. No software tool, regardless of its vehicle support coverage or technical sophistication, can retrieve data from a module that cannot communicate. Until the methodology presented in this paper, no documented solution existed for recovering EDR crash data from a completely non-communicative ACM that had sustained this level of physical damage. The present case, therefore, represents not only a forensic milestone for the specific accident under investigation, but a methodological breakthrough with broader implications for the field of forensic accident reconstruction.

The module in question — a Bosch SRS ACM from a 2021 Mercedes-Benz Sprinter. The vehicle's ACM was non-functional, and standard diagnostic extraction using CrashScan had confirmed that no communication could be established. Rather than concluding that the EDR data was irrecoverable, the authors initiated a hardware-level forensic investigation, the results of which form the subject of this paper. With the donor unit containing the recovered data, standard bench-top EDR extraction was performed using Collision Sciences CrashScan, yielding a full EDR report. The report documents two recorded crash events: a historical 1st Prior Event (ignition cycle 21,877) with a peak longitudinal delta-V of −2 km/h, and the forensically relevant Most Recent Event (ignition cycle 26,577) corresponding to the accident under investigation, with a peak longitudinal delta-V of −20 km/h. Pre-crash data confirms that the vehicle was braking intensively for more than 4.5 seconds prior to impact, with ABS and ESC actively engaged. No airbags were deployed. The forensic significance of the recovered data, the technical novelty of the methodology, and the complete accident reconstruction are presented and discussed. The consistency and reliability of the data acquired using CrashScan have been repeatedly validated in previous studies involving Mercedes-Benz vehicles through comprehensive comparative analyses against data retrieved with the Xentry diagnostic platform [8,9].

This paper is organised as follows. Section 2 gives brief introduction to data extraction methods. Section 3 presents the initial observation of damaged SRS module, and Section 4 describes the structural analysis of its memory chip. Data extraction process is outlined in Section 5 and EDR extraction in Section 6. This is followed by discussion in Section 7 and conclusion in Section 8.

## II. Background

A typical structure of the path from OBD2 down to the bits of information stored inside Flash EEPROM chip is presented in Figure 1. Different approach should be taken for establishing access to SRS data at different points.

### A. CAN Diagnostic Interface

Standard tools for scanning codes via OBD2 connector do not support access to SRS module. This will require some specialised tools for accessing data inside SRS module via standard OBD2 connector. For example, CrashScan Lab Kit from Collision Sciences to retrieve EDR data [10]. Alternatively, someone could reverse engineer the SRS module and analyse its firmware to find all the necessary commands and protocols [3].

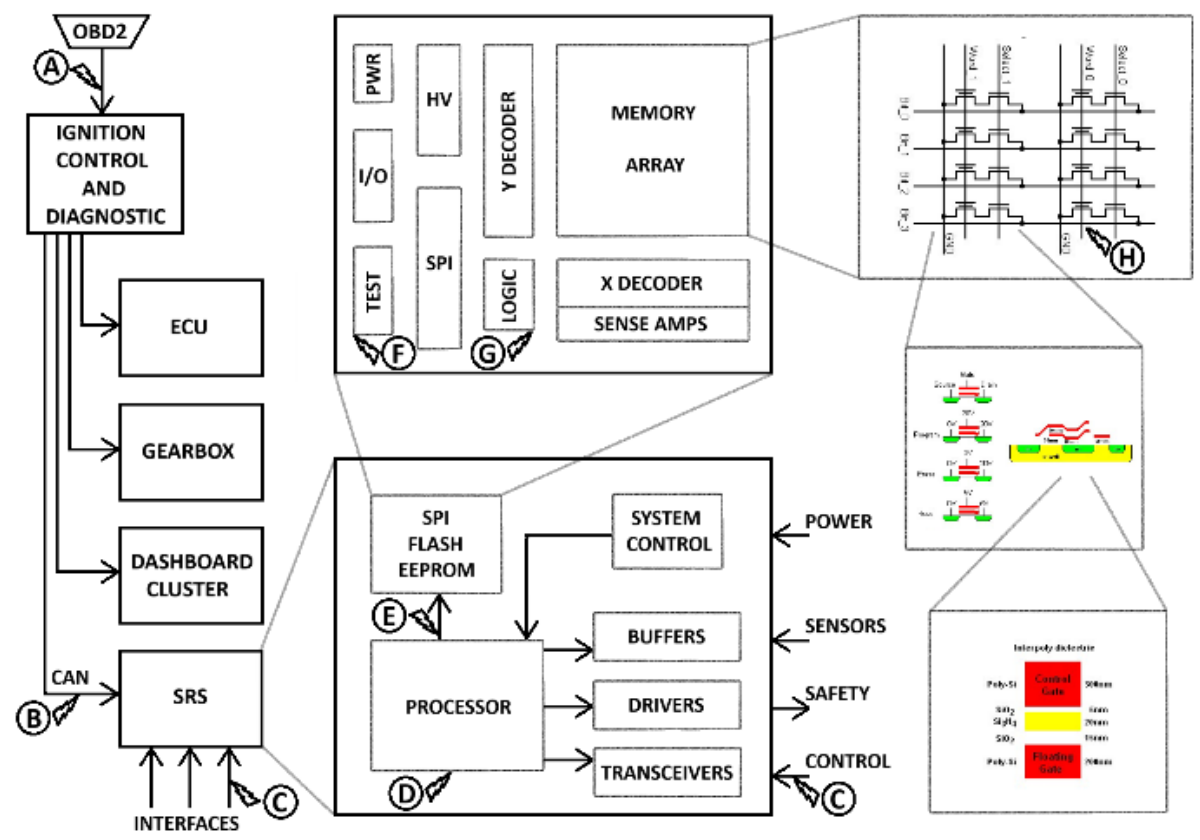

Fig. 1. The long path from OBD2 to stored information

### B. Local CAN Interface

If the access via OBD2 is not possible because of the damage to other modules or the CAN network, the SRS module can be accessed directly through the same CAN network. This might not even require the physical removal of the module but just unplugging one of its connectors and plugging a diagnostic tool. However, as the communication will be going directly thus bypassing the control module, the protocol and commands could be different. Therefore, some sort of preliminary work will be required to understand the protocol. For example, by analysing such communication during diagnostic tool operation or through reverse engineering of the SRS module.

### C. Auxiliary Interfaces

There are other interfaces usually present on the SRS module apart from the main CAN. These could be other CAN networks, LIN and other interfaces. Still, this might be done without physical removal of the module but just unplugging one of its connectors and plugging a diagnostic tool. Very likely this would require detailed understanding of the module operation by means of reverse engineering.

### D. SRS Processor

The processor chip inside SRS module has full control over the data stored inside the memory chip. Hence, once the access to it is established all the stored information could be retrieved and subsequently analysed. However, this will require physical removal of the SRS module from the car and accessing it in a lab environment. More detailed analysis of the firmware stored inside the processor will be required to find the control points and establish a connection via test points usually present on a PCB. Firmware extraction itself might also require sophisticated methods for approach [11]. These methods can be split into several categories. Non-invasive which are usually low-cost and involve observations of the device operation or manipulations with external signals. They require only moderately sophisticated equipment and knowledge to implement. Invasive methods, in contrast, are expensive and require sophisticated equipment and knowledgeable engineers. However, they offer almost unlimited capabilities to extract information from chips and understand their functionality. Semi-invasive methods fill the gap between non-invasive and invasive, but usually more affordable. For these the chip needs to be de-packaged but the internal structure remains intact. In most cases the chip remains fully operational.

Tools required for non-invasive methods usually available at most electronics engineering labs. These tools involve digital multimeter, IC soldering/desoldering station, universal programmer, oscilloscope, logic analyser, signal generator, power supply, PC and prototyping boards. Non-invasive methods can be divided into side-channel (timing [12], power analysis [13], emission analysis [14]), data remanence [15], data mirroring [16], fault injection (glitching [17], bumping [18]) and brute forcing [19]. Tools used for invasive methods involve simple chemical lab, high-resolution optical microscope, wire bonding machine, laser cutting system, micro-probing station, oscilloscope, logic analyser, signal generator, power supply, scanning electron microscope (SEM) and focus ion beam (FIB) workstation. Invasive methods can be divided into sample preparation [20], imaging [21], direct memory extraction [22], reverse engineering [23], micro-probing [24], fault injection [25] and chip modification [26]. Tools used for semi-invasive methods involve simple chemical lab, high-resolution optical microscope, UV light source, lasers, oscilloscope, logic analyser, signal generator, PC and prototyping boards. Semi-invasive methods can be divided into imaging [27], laser scanning [28], optical fault injection [29], optical emission analysis [30] and combined methods [31].

### E. *Memory Interface*

After desoldering the memory chip from PCB, it could be read by various tools. In some cases, the memory chip can be read without removing it from the PCB, for example, by using test clips with universal programmer supporting SPI communication via ISP interface. Most memory chips have full datasheet with detailed description of their interface and protocol for read/write access. Anyone with basic electronics and programming skills would be able to read the information stored inside a memory chip. However, the data structure of the file extracted from the chip might not be the same as read by dedicated tools via OBD2 or another interface. Therefore, some kind of reverse engineering is likely to be necessary to decode the raw data from the file created by reading the memory chip.

### F. *Factory Debug Access*

Many chips have dedicated factory test interfaces. They might have already present on some additional pins or share their functionality with existing I/O pins, for example, when a specific voltage is applied to a dedicated Test pin. This feature is usually used for quick testing of the internal chip functionality on the production line. However, these interfaces could also be used for alternative access to the internal chip memory arrays in case the main access interface is damaged or not functioning [32,33]. The successful usage of this interface will often require full chip reverse engineering to find the communication protocol [11]. This is often time consuming and expensive process [34].

### G. *On-chip Test Points*

Many silicon chips have dedicated factory test points on their surface. These are normally used at chip factory to perform failure analysis testing in case something goes wrong during the manufacturing process. But the same points could be used for low-level access to the memory storage array. Very often these test points are covered by passivation protection layer to prevent any damages to the die surface from the environment. Therefore, any access to these points will require the use of a laser cutter [14] or FIB tool [26]. Moreover, the successful usage of this approach will often require full chip reverse engineering to find the functionality of each test point. This is often time consuming and expensive process.

### H. *Direct Memory Imaging*

This approach might be the only choice if the memory chip is severely damaged electrically, mechanically or by fire. It does not require the knowledge of the chip access interface, but it requires the knowledge about the data bits and addresses allocation inside the memory array [3,22,35].

## III. Components Level Analysis

The pictures of the PCB from damaged SRS module are presented in Figures 2 and 3.

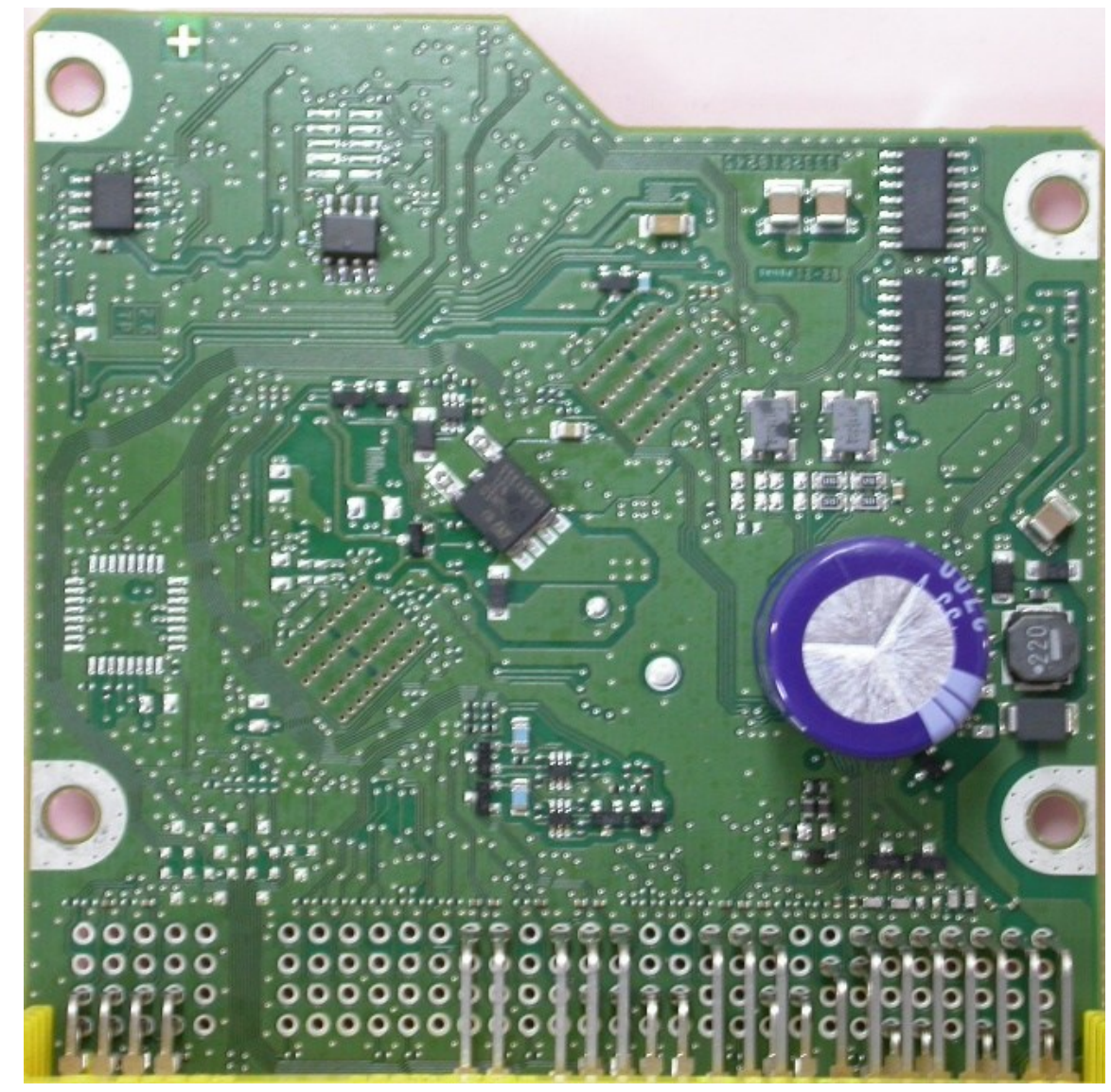

Fig. 2. Top side of the SRS module PCB

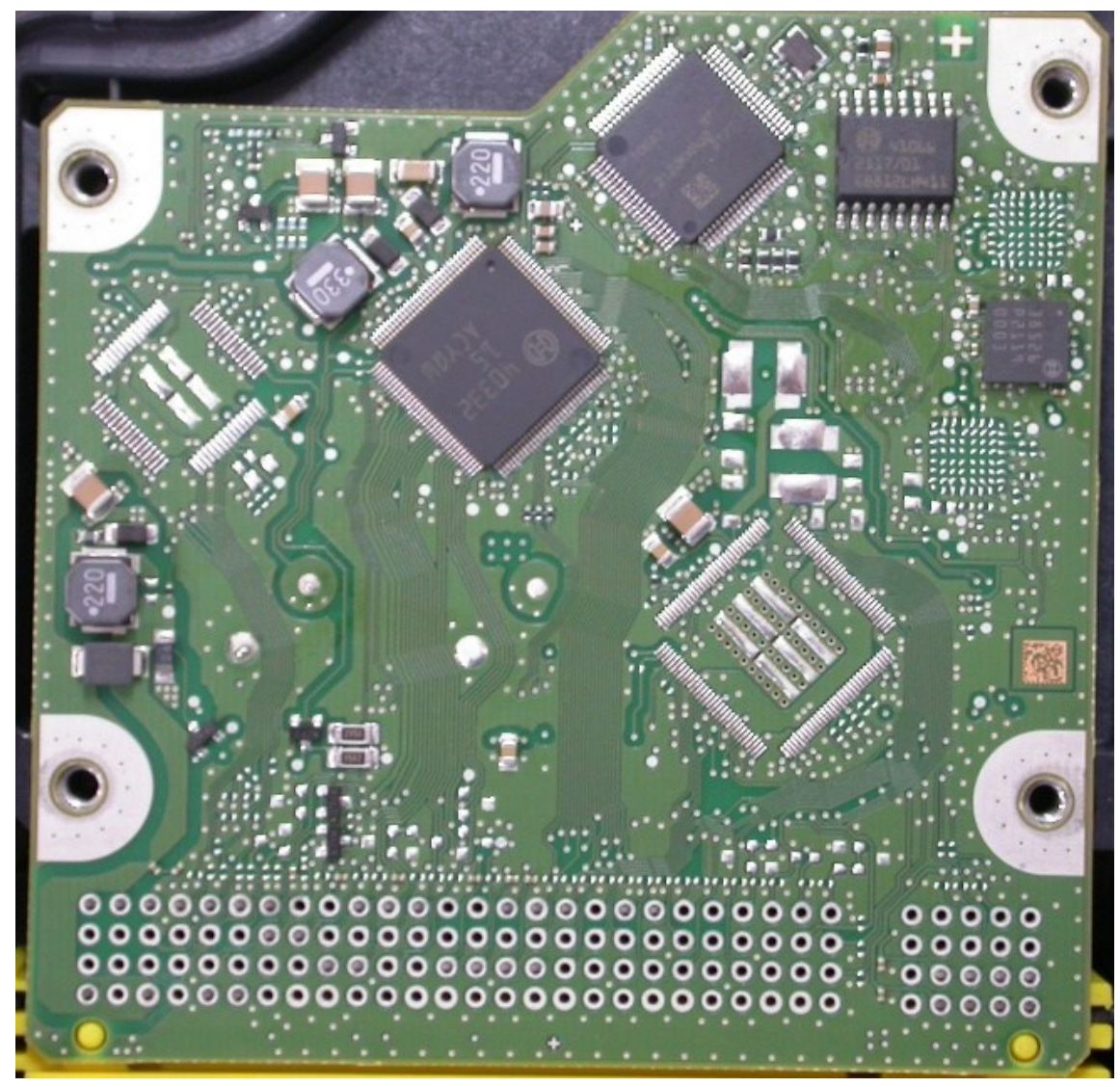

Fig. 3. Rear side of the SRS module PCB

Two components on the PCB have noticeable damage:

- SPI EEPROM in SOIC8 marked 95128RT K114K
- Processor in TQFP80 marked 10643 2122KXR02

According to its marking, the memory chip is M95128 from STMicroelectronics with SPI bus and 16 KB of EEPROM memory [36]. The processor chip has custom factory marking and cannot be directly linked to any standard microcontroller at this stage. However, the marking on its silicon die could reveal the actual name of the device.

Both devices were initially removed with hot air gun and measured for electrical damages using multimeter to evaluate the level of the damage to their pins. The following pins were found damaged on the memory chip:

- pin 2 (Q) has low resistance to pin 4 (Vss)
- pin 6 (C) has no connection
- pin 7 (H) has low resistance to pin 4 (Vss)
- pin 8 (Vcc) has low resistance to pin 4 (Vss)

The memory chip has the following connections to the processor chip:

- pin 1 (CS) to pin 14
- pin 2 (Q) to pin 11
- pin 5 (D) to pin 12
- pin 6 (C) to pin 13
- pin 7 (H) to pin 72
- pin 8 (Vcc) to pin 10

The memory chip was partially decapsulated to observe the damages. The result is presented in Figure 4. The image of the chip surface is presented in Figure 5. The silicon die has the following markings on it: ST 2012 M95128KA.

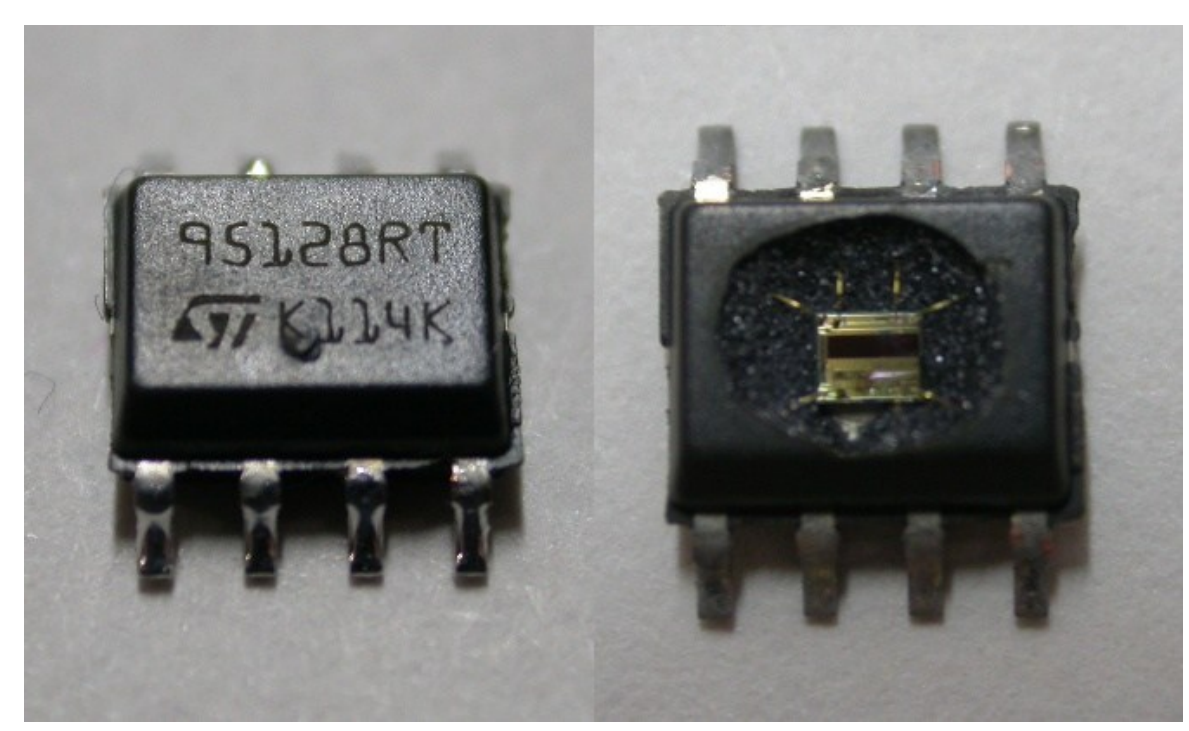

Fig. 4. Images of chip: a) desoldered; b) partially decapsulated

Figure 6 shows images of the damaged pins at higher magnification.

It can be observed that pin 2 has very severe damage with bonding wire detached and surface area burned. Pins 6, 7 and 8 has some noticeable damages and this affected their functionality as well. Very likely the chip has sustained high current surge. However, this did not result in any severe structural damage or cracks.

Although there is no structural damage to the EEPROM memory array located in the middle of the silicon die, there is no guarantee that the data inside it remained intact. This is because the cause of this damage is unknown. However, the only practical way to extract the data from the memory array of such severely damaged chip would be by direct imaging of the electrical charge in all memory cells of the EEPROM array.

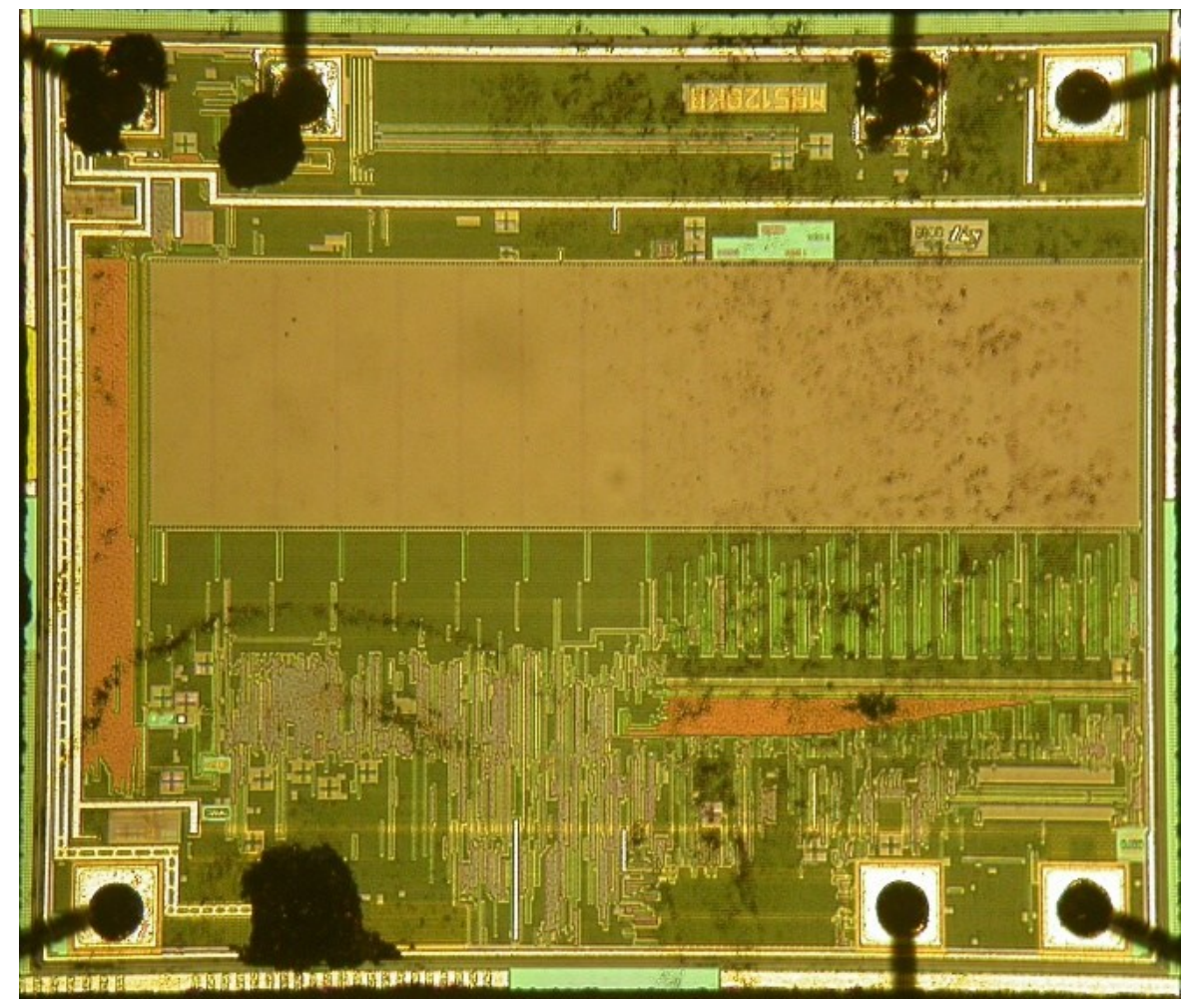

Fig. 5. Chip surface after partial decapsulation

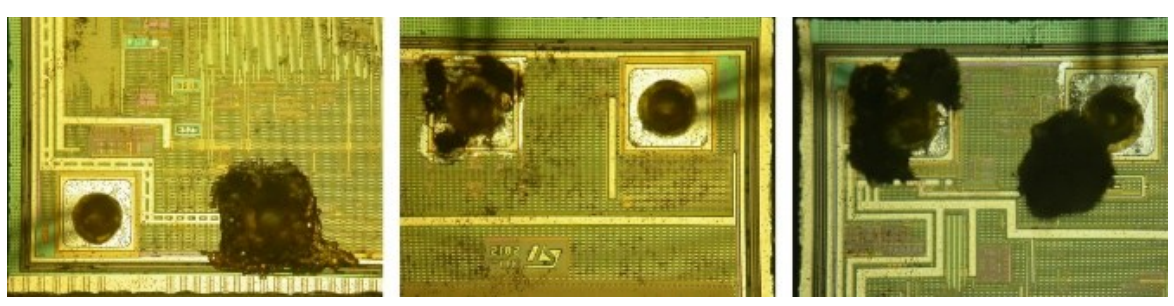

Fig. 6. Images of damages: a) pins 1 & 2; b) pins 5 & 6; c) pins 7 & 8

Then the processor chip was partially decapsulated to observe the damages. The result is presented in Figure 7. The image of the chip surface with surrounding bonding wires is presented in Figure 8. The wires near damaged pins have their pin numbers marked for easier navigation.

The silicon die has the following markings on it: RENESAS R7F7 01025. This corresponds to a standard Renesas microcontroller from RH850/F1L family [37]. This die covers several products up to 1 MB of Flash and up to 100 pins packages. Because of 80-pin package its real order number should be R7F7010203AFP or R7F7010204AFP.

High magnification images of the areas around the damaged pins are presented in Figure 9. It can be observed that the bonding wire to pin 10 (EVCC) and one of the wires to pin 27 (REGVCC) are darker. This indicates a high current flow through these wires resulting in overheating and plastic decomposition. Pins 10, 11, 13, 14, 15, 16, 17, 27, 42, 72 and 76 has some noticeable damages and this very likely affected their functionality.

The following pins of the processor were previously found damaged using multimeter:

- pin 10 (EVCC) has lower than usual resistance to pin 16 (EVSS)
- pin 11 (P0_4) has no connection
- pin 13 (P0_6) is short circuited to pin 14 (P0_11)
- pin 15 (P0_12) has low resistance to pin 16 (EVSS)
- pin 72 (P11_1) has no connection
- pin 76 (EVCC) has no connection

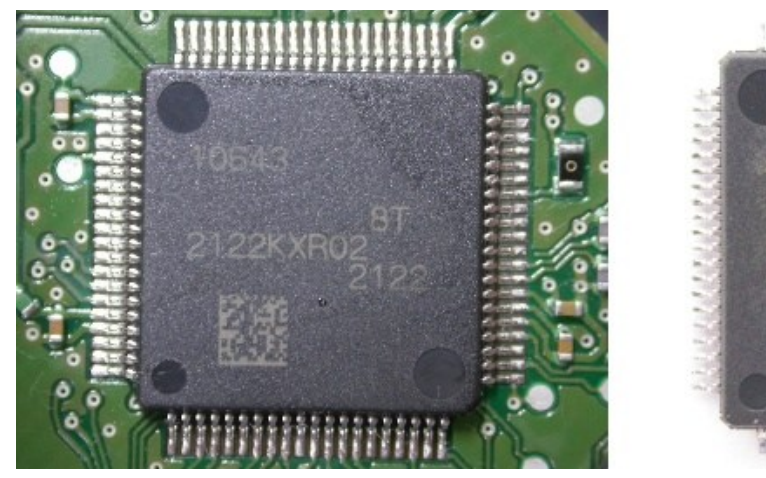

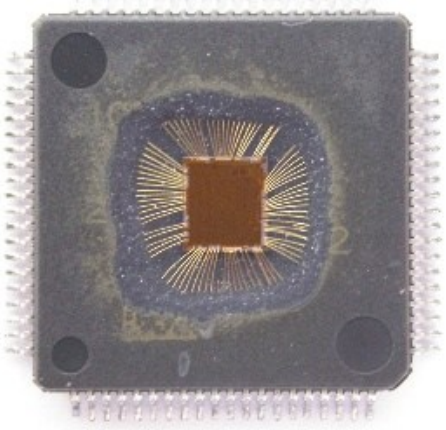

Fig. 7. Images of chip: a) on PCB; b) partially decapsulated

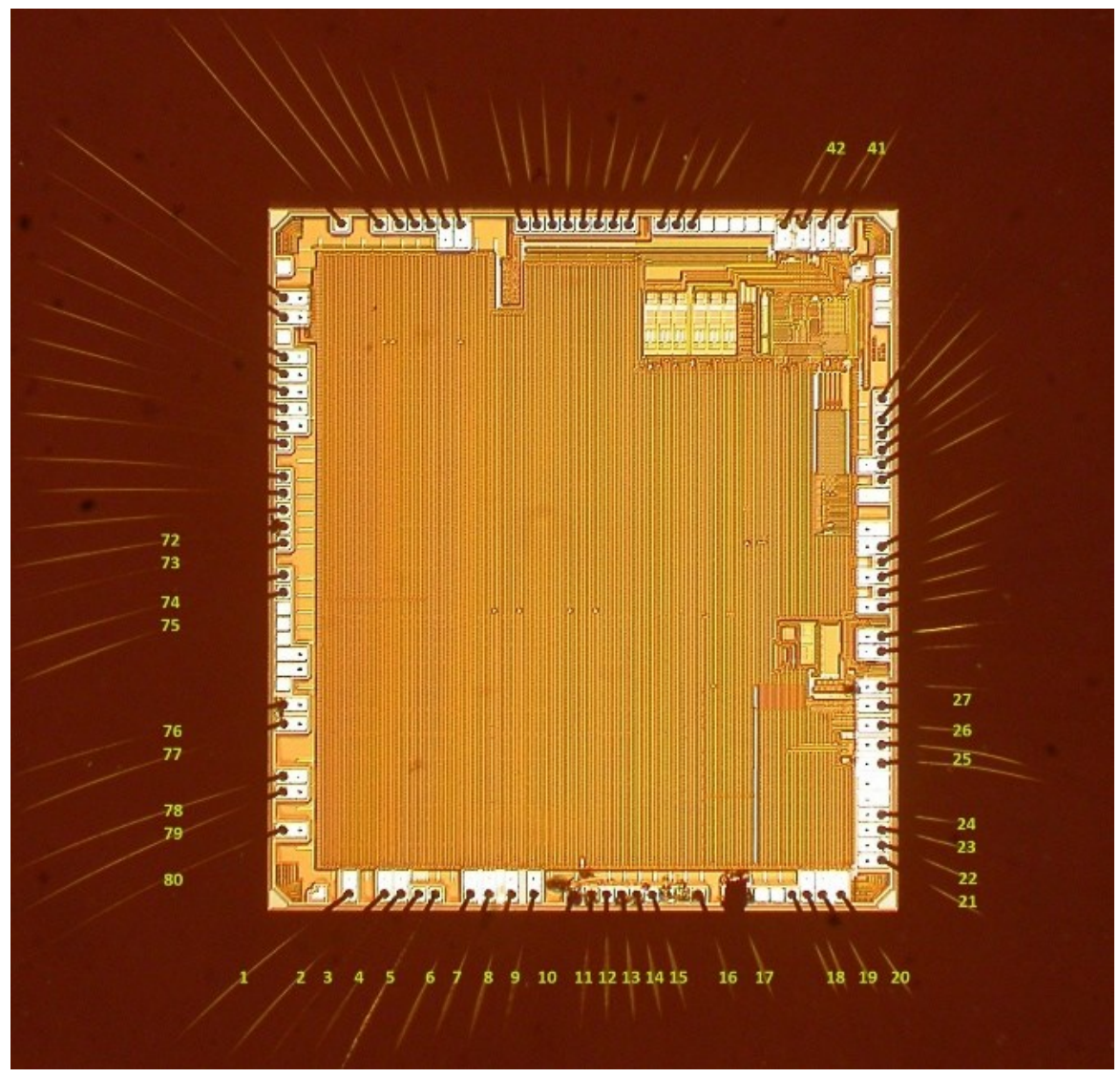

Fig. 8. Chip surface with bonding wires after partial decapsulation

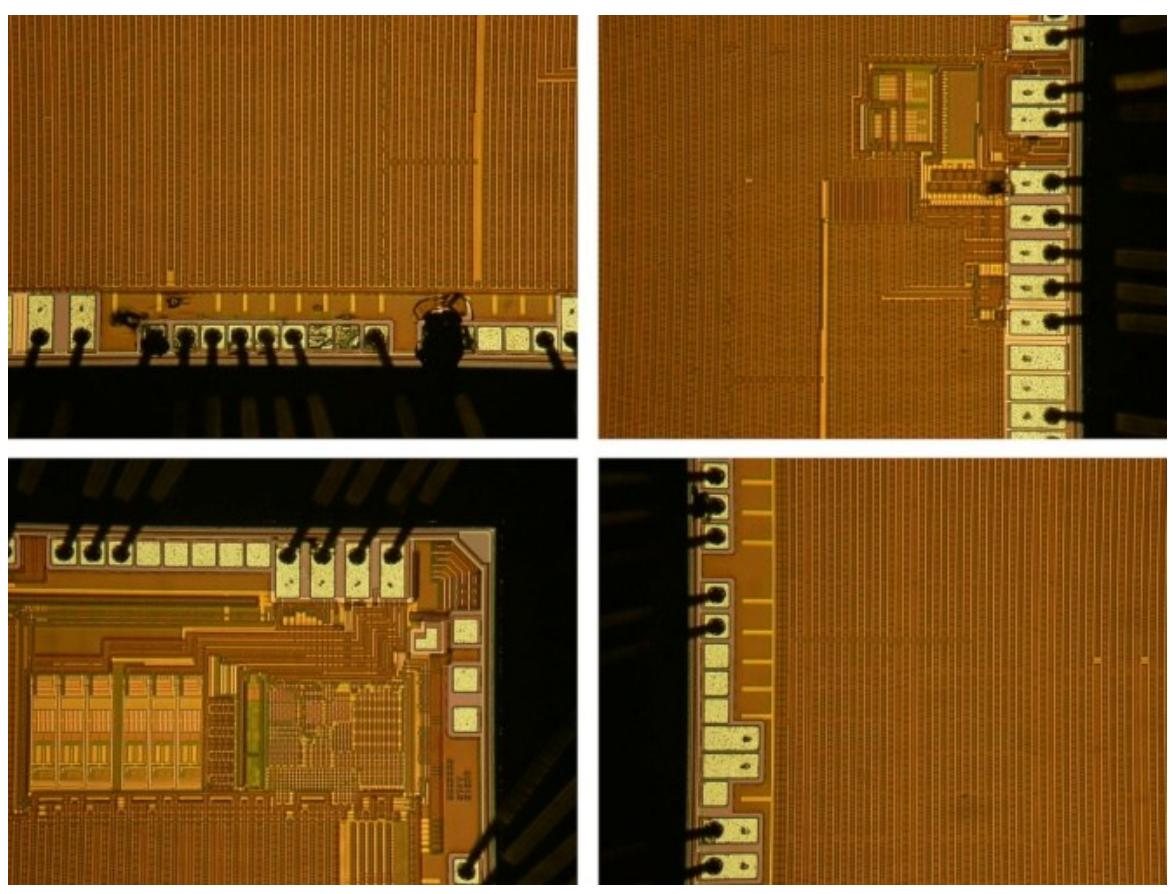

Fig. 9. Images of damages near pins: a) 10…17; b) 27; c) 42; d) 72, 76

Although the actual cause of the damage is unknown very likely it was caused by the increased current flow from power supply pins to GND and some other pins. Because on the PCB the power is supplied simultaneously to pins 10, 27, 42 and 76, all these pins were damaged as a result. The I/O pins connected to SPI EEPROM could have been damaged from the processor side by the surge in the EVCC line, because these pins (11, 12, 13, 14 and 15) are located in between EVCC and EVSS. Alternatively, the surge in the power supply line could have damaged the memory chip first and then the I/O pins of the processor connected to SPI chip were damaged. However, none of the above scenarios could explain why pin 17 on the processor was also damaged. Because this pin is not wired to any active component on the PCB — only to unpopulated QFP64 device. But this could be the outcome of one possible scenario: electromagnetic pulse induced high voltage on this floating pin 17 and caused a latch-up resulting in short circuiting power supply to ground. This overloaded the power supply IC resulting in a very high current flow. When this current overheated the wires and caused significant damage to the area between pins 10 and 16, the supply was interrupted by a burned wire. This immediately caused overrun in the power supply IC resulting in a high voltage spike up to the input of above +12 V. This second wave caused the actual damage to the SPI EEPROM because its maximum supply voltage is 6.5 V. In addition, this caused further damage to the processor.

## IV. Chip Level Analysis

To avoid damaging the original SPI EEPROM with data, some blank samples were tested first. That helped in choosing the best sample preparation methodology. To start with, the same type of the blank samples as the original device had to be ordered. However, there was no information about the package markings in all the datasheets on M95128 devices. Hence, different types of M95128 chips were ordered from electronics distributor. The pictures of the devices and their respective order codes are presented in Figure 10. This helped to determine the full name of the original device as M95128-DRMN3 and to confirm its correct datasheet [36].

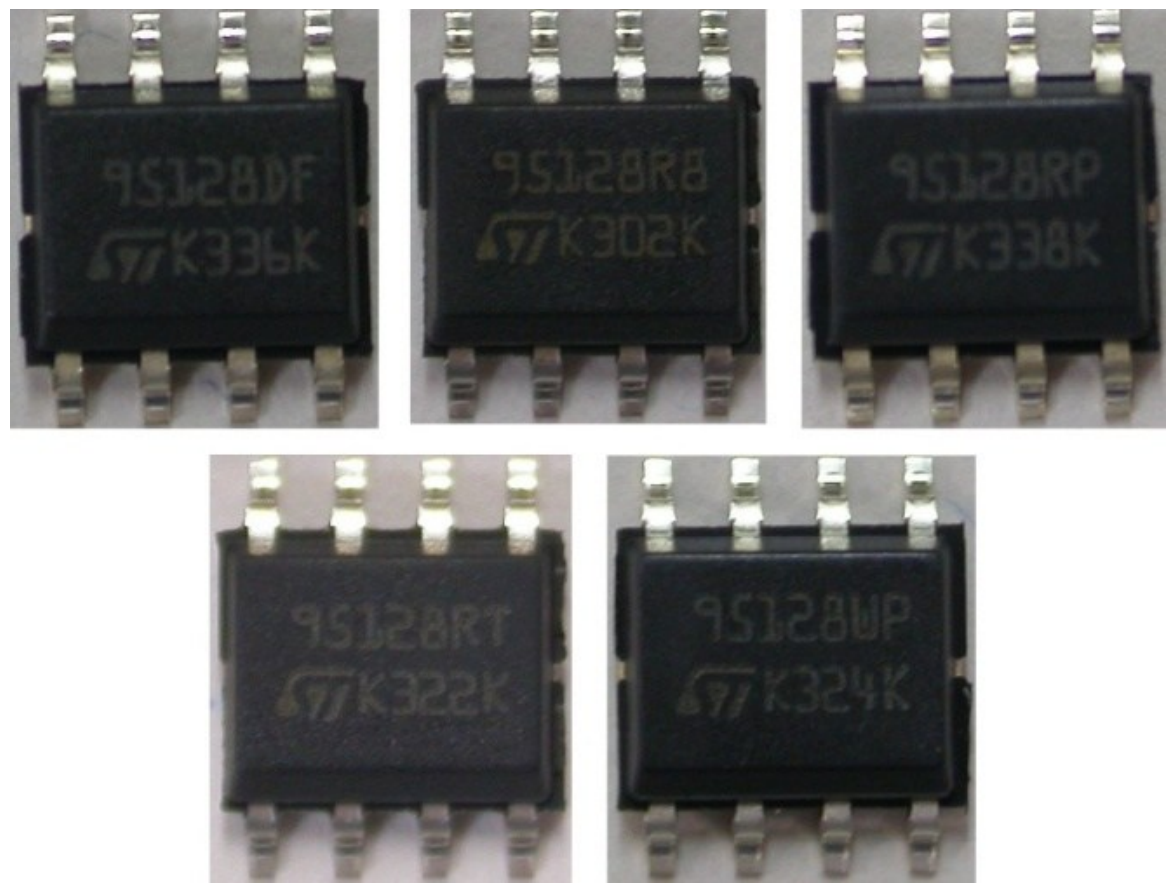

Fig. 10. Images of M95128 in SOIC8 with suffixes: a) DFMN6; b) DRMN8; c) RMN6; d) DRMN3; e) WMN6

The result of the package analysis of a blank M95128-DRMN3 device is presented in Figure 11. Although there could be some variations in the thicknesses, they are usually within 10 µm tolerance for metal/glue/silicon and within 50 µm tolerance for plastic.

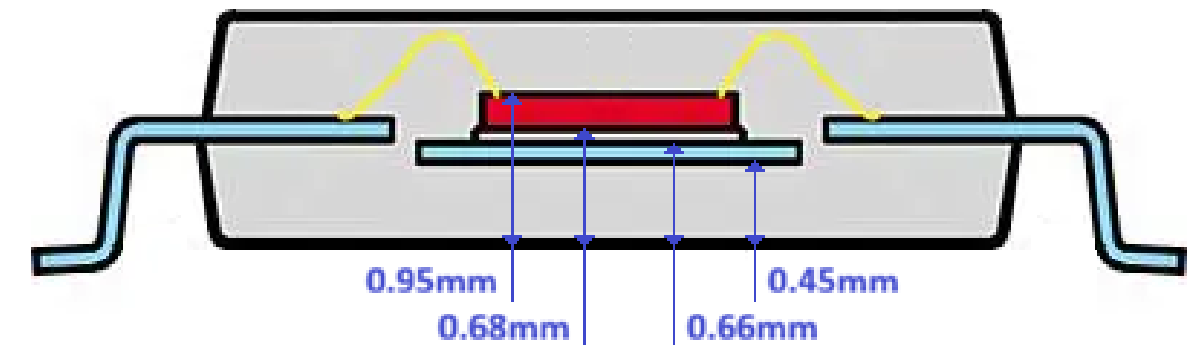

Fig. 11. Internal sizes of M95128-DRMN3 device

The knowledge of the internal device structure helps in setting the correct parameters during sample preparation process and minimizes the risk of damaging the sample. The overall structure of the device can be found in the datasheet on page 3 [36].

Two blank samples were programmed with different test data to reveal the physical address allocation and error correction encoding. Both samples went through full sample preparation procedure. Then they were fully imaged and digitised. By analysing the correlation between the cell charge images and the programmed data it was possible to build a map of the EEPROM memory array. This map annotated allocations for individual data bits in each block, allocation of addresses along memory rows and allocation of addresses along memory columns. This helps with the correct conversion from digitised image of the memory array into HEX file.

According to the datasheet page 21 [36], the EEPROM array in M95128 device has error correction feature — ECC. It helps in reducing the data errors by correcting single bit failures. However, it can also help in confirming that full data extraction was carried out without errors. Ideally, all ECC codes must match for 100% error free extraction. The polynomials for this ECC function must be mathematically extracted through data analysis of the blank samples. For this memory device there are 6 error correction bits for every 32 bits of data. The required 6 polynomials were reconstructed by processing 32 discrete mathematical equations.

## V. DATA EXTRACTION

After carrying out successful sample preparation of a blank M95128-DRMN3 device, the memory chip from the SRS module was prepared in the same way. The result of the polishing process starting from exposing the copper heatsink carrier down to 10 µm of remaining bulk silicon substrate is presented in Figure 12.

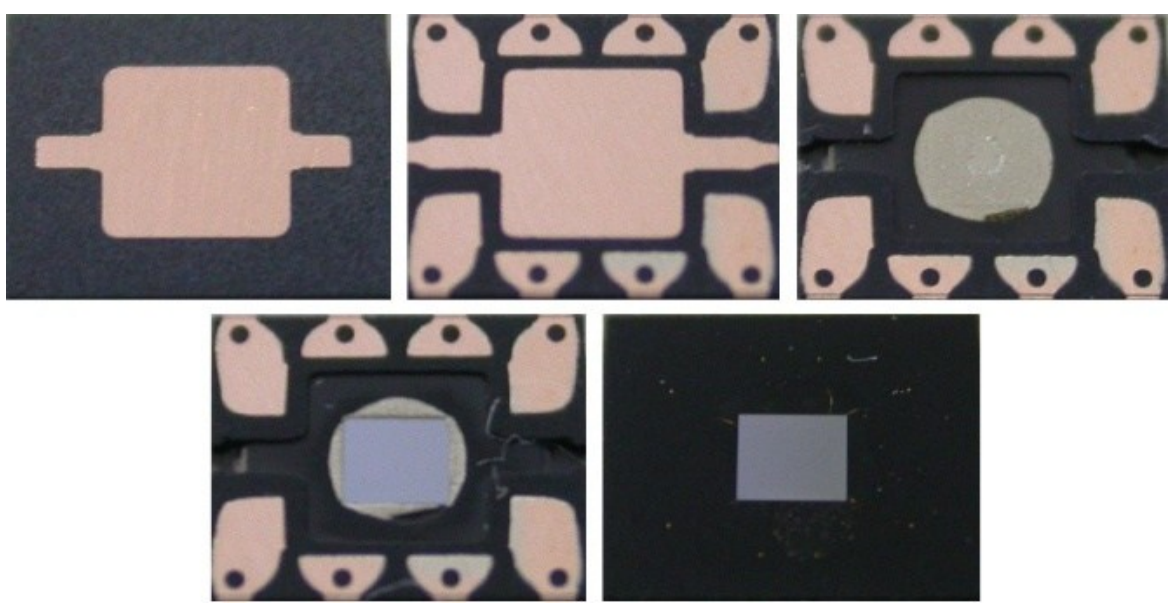

Fig. 12. Polishing stages: a) carrier; b) leads; c) glue; d) substrate; e) 10 µm

When the thickness of silicon substrate was reduced to 50 µm, it became possible to observe the internal structure using NIR camera. The damages to the pins 2, 6, 7 and 8 were imaged (Figure 13) for later comparison with fully removed silicon observation.

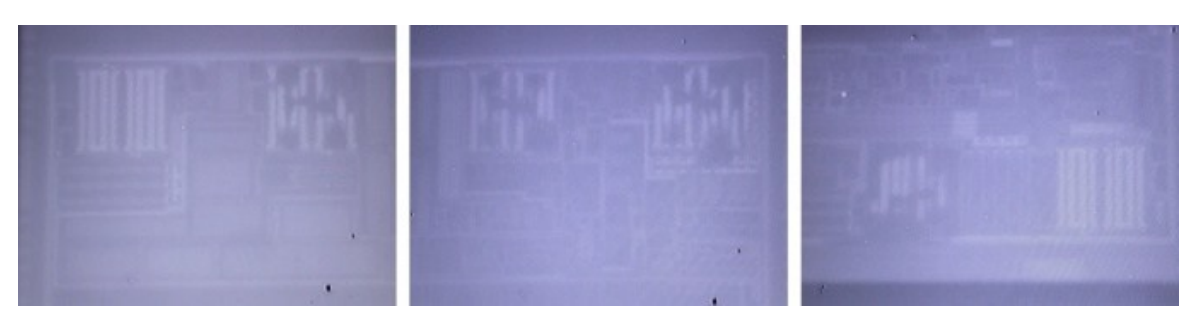

Fig. 13. NIR backside images: a) pins 1 & 2; b) pins 5 & 6; c) pins 7 & 8

At the substrate thickness of 5 µm it was possible to see the damaged area near pin 2. This means that the damage to the pin 2 was substantial and caused not only the metal layers to burn but also melted and evaporated part of the silicon substrate. The observation of this damage is presented in Figure 14. The cavity was filled by melted gold from the nearby bonding wire.

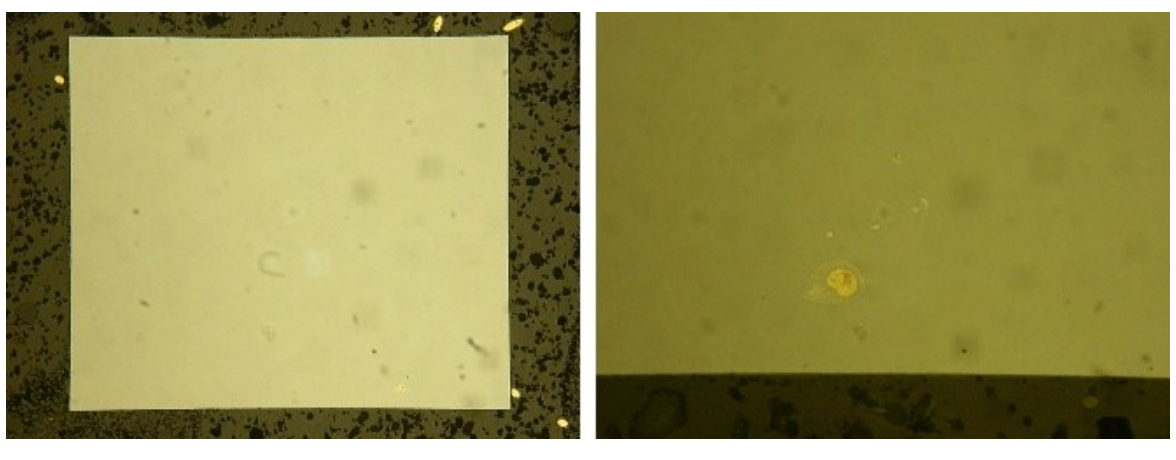

Fig. 14. Damages observed at 5 µm thickness: a) whole chip; b) near pin 2

The major concern during the removal of the remaining silicon substrate was about the mechanical strength of the active chip areas. Fortunately, the severe damage near pin 2 did not affect other chip structures. The result of the chemical removal of the remaining silicon is presented in Figure 15. All the damages to the pins 2, 6, 7 and 8 areas are clearly visible (Figure 16). However, no damage can be seen around the EEPROM memory array. At this point, the original device is ready for imaging of the electrical charges inside EEPROM memory cells. However, to be on a safe side, some blank samples were prepared and tested first to avoid the loss of important data in the real device.

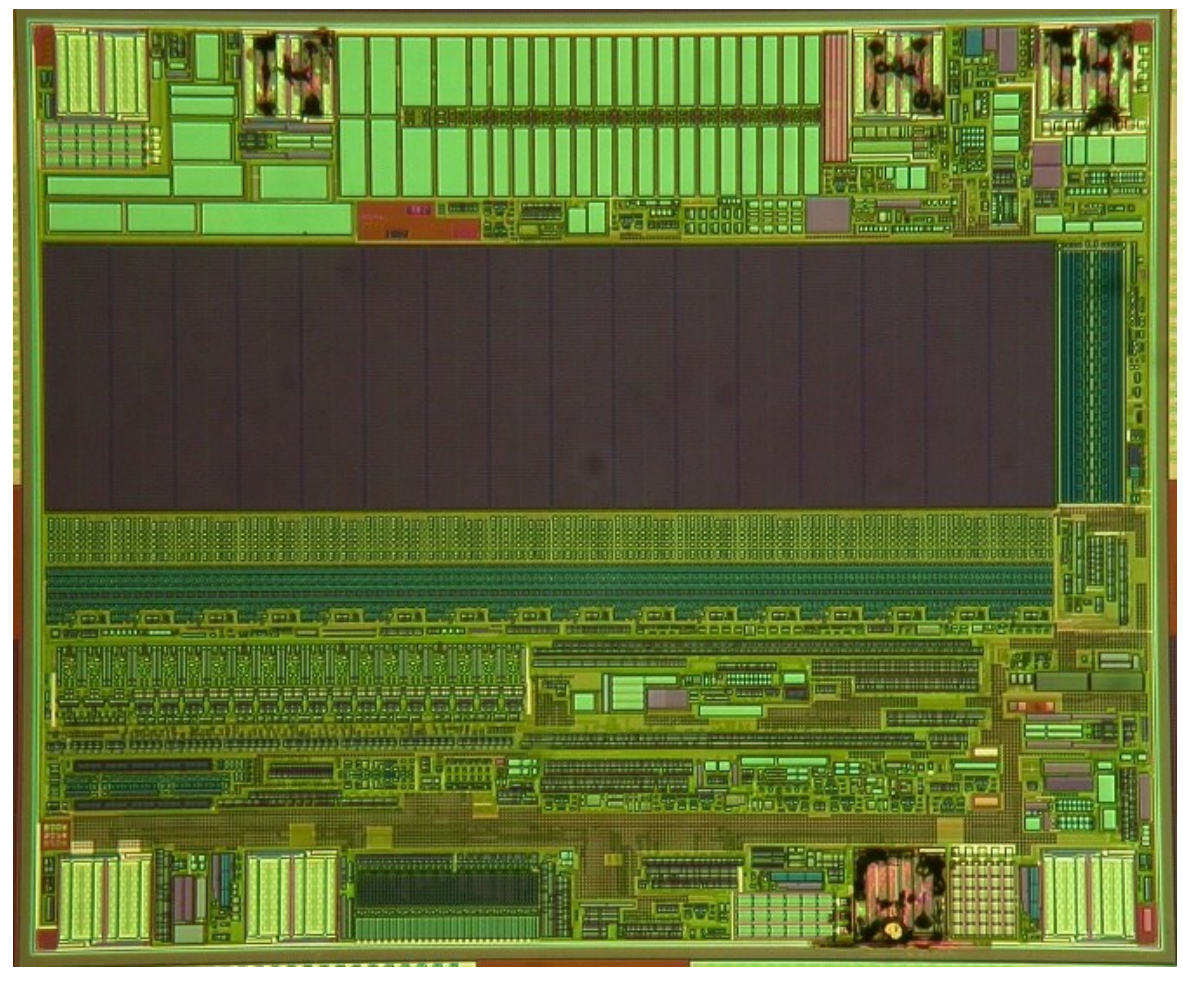

Fig. 15. Result of sample preparation with silicon die view

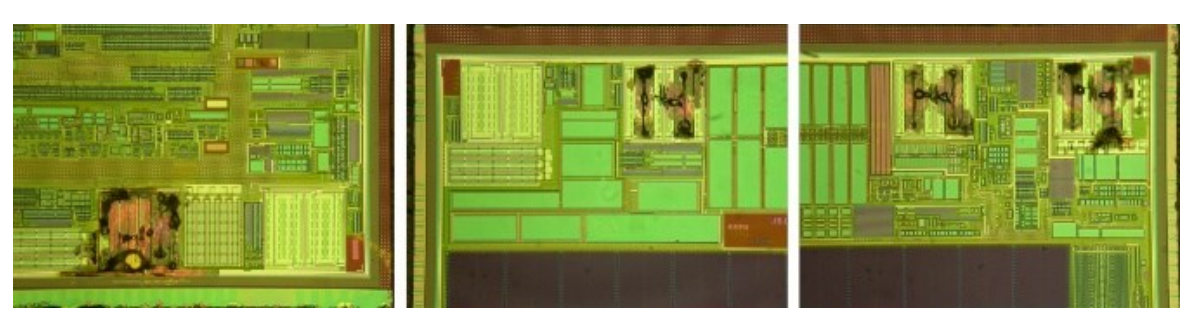

Fig. 16. Images of damages: a) pins 1 & 2; b) pins 5 & 6; c) pins 7 & 8

Initially, some extra imaging tests on the blank samples were carried out to determine the correct and safe parameters. The result of the brief testing of the original device is presented in Figure 17. These areas are taken at corners to minimise any disturbance to the remaining data. The testing was carried out with minimal damage to the original device; therefore, the quality of the images is far from perfect, and a lot of noise is present. Still, it can be observed that the data blocks consist of 32 data bits (dark in blank areas) and 6 error correction bits (bright in blank areas). From the above observations it can be concluded that the internal EEPROM data survived the electrical damage to the memory chip.

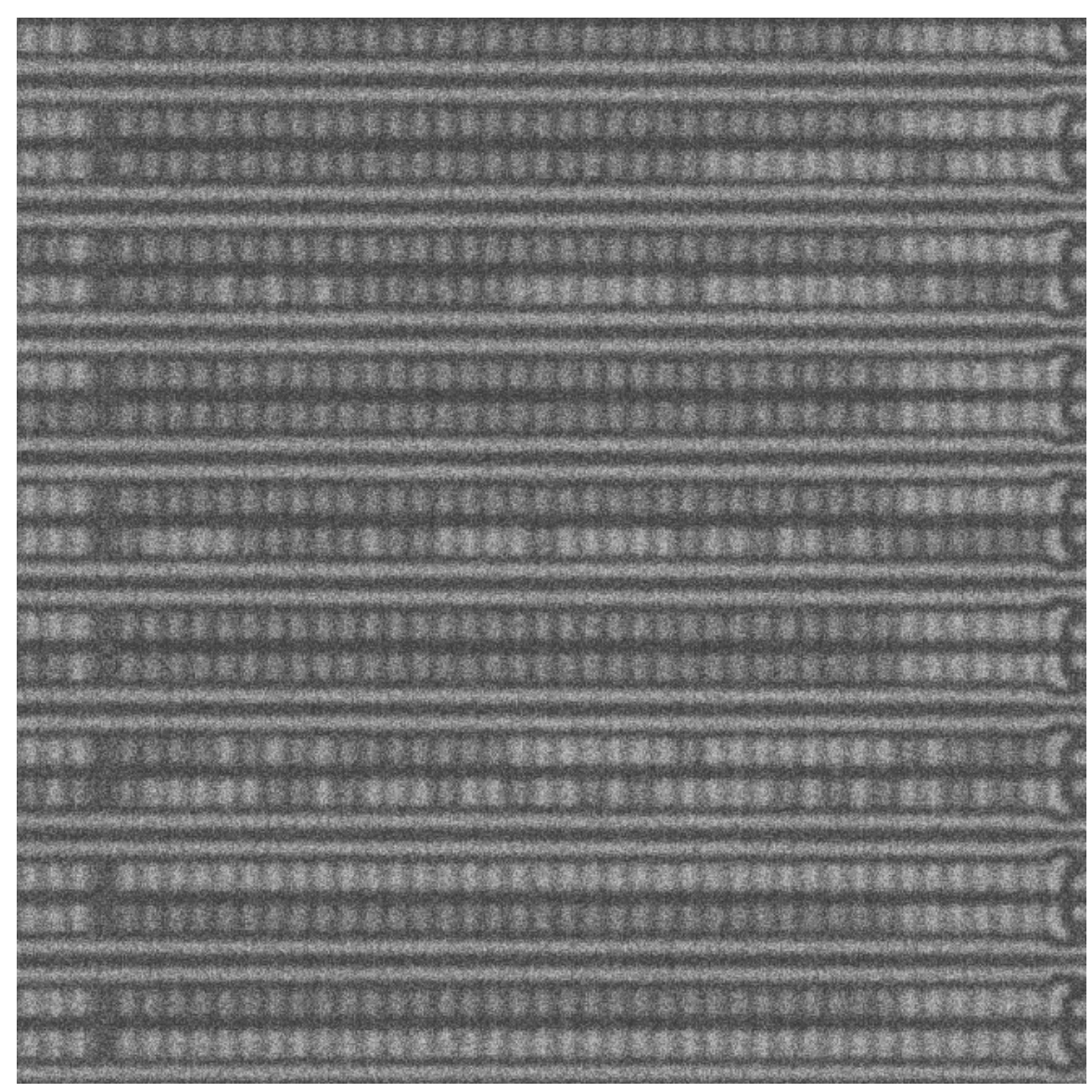

Fig. 17. EEPROM array cells charge image

After successful testing of the blank samples, it became possible to extract all the data from the damaged SPI EEPROM device from the SRS module. To achieve this, the whole image of the memory array was first digitised into a bitmap. Then this bitmap was verified against error correction bits, and no single error was found. After that the data bits were combined according to the physical map of addresses to produce the resulting HEX file.

There is no guarantee that all the data inside the memory chip survived the device damage until the data is analysed. This is because the cause of this damage is unknown. If this happened during the crash, then everything depends on the exact moment of this. If this happened after the crash, then the data are likely to be preserved. The only thing that remained uncertain at this stage is whether all the data in the EEPROM memory array survived the electrical damage of the chip.

Two approaches were successfully used to recover the actual crash data. The first involved using a donation SRS module with same part number from the same car model. Its SPI EEPROM chip was removed and reprogrammed with the extracted HEX file, then soldered back. In this case a standard SRS data analysis software was able to read all the data correctly. Another approach involved reverse engineering of the data structure inside the SPI EEPROM. Although this was more tedious and time-consuming process, it allowed better understanding of the information stored inside the memory chip and allowed its exact interpretation.

## VI. EDR EXTRACTION

With the complete binary content of the original EEPROM successfully extracted as a verified HEX file, the next challenge was to reconstruct a functional ACM unit capable of communicating the recovered data through a standard diagnostic interface. This was achieved by identifying a compatible donor ACM unit — a Bosch SRS module of the same hardware and software version as the original — and transferring the recovered binary data to its EEPROM.

It must be emphasized that this step required highly specialised technical knowledge extending well beyond standard forensic EDR practice. The binary data recovered by SEM imaging reflects the raw memory contents of the EEPROM as stored by the original module's firmware. The M95128 EEPROM uses a proprietary binary encoding scheme defined by the Bosch SRS firmware, encompassing not only the EDR data records but also module configuration data, calibration parameters, security access credentials, and system state information. A naive byte-for-byte copy of the data to a donor unit EEPROM would in most cases render the donor unit non-functional, as critical parameters such as VIN coding, immobilizer data, and module-specific calibration values would be inconsistent with the donor unit's hardware identity.

The successful data transfer required expert-level understanding of the Bosch SRS EEPROM memory map — identifying and preserving the forensically relevant crash event records and EDR data structures while appropriately handling module-specific configuration parameters that are not subject to forensic transfer. This process required advanced knowledge of EEPROM binary encoding conventions, SPI communication protocols, ACM firmware architecture, and the specific data layout of the Bosch 0 285 015 383 module family. The technical details of this procedure constitute specialist forensic know-how that is documented in the case record but are not disclosed in full in this publication, in accordance with standard forensic practice for sensitive methodology.

Following the data transfer, the donor ACM unit was verified for basic electrical functionality and prepared for bench-top diagnostic extraction.

### *A. Bench-Top EDR extraction using CrashScan*

EDR data extraction from the donor ACM was performed using the bench-top methodology described in detail in prior publications [6,8]. The donor ACM was removed from any vehicle environment and connected to the CrashScan EDR Lab Kit on the laboratory bench (Figure 18). The CrashScan OBD Breakout Cable provides the necessary CAN-H, CAN-L, ground, and +12 V supply connections to the module without requiring vehicle-specific harnesses (Figure 19). Network termination was achieved with a 60 Ω resistor appropriate for the module's CAN bus topology.

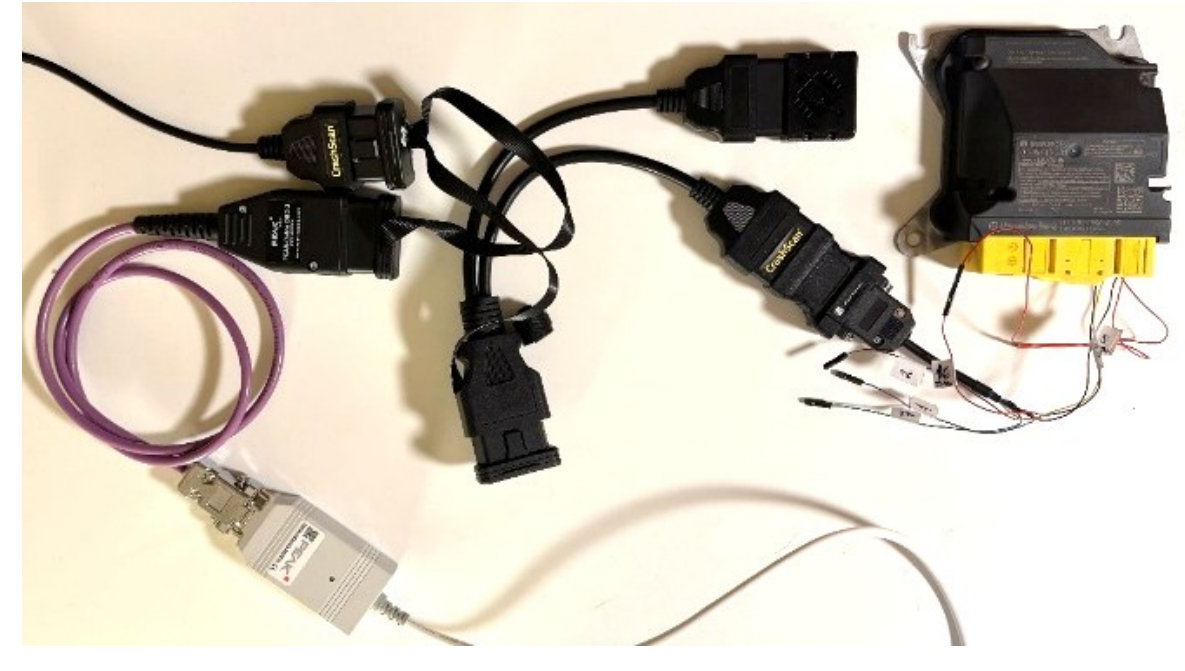

Fig. 18. Connecting the CrashScan Lab Kit to the ACM

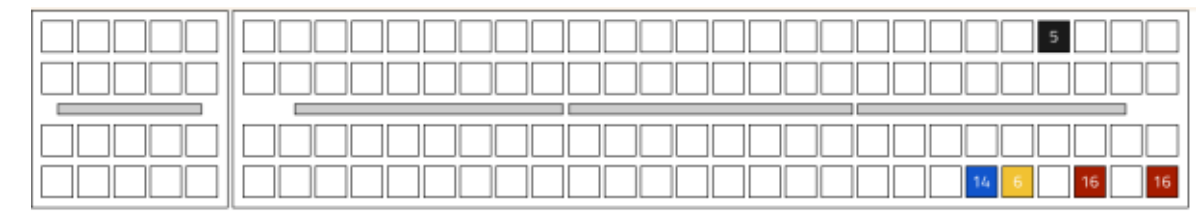

Fig. 19. Wiring diagram for CrashScan Lab Kit (CAN H and L, GND, +12V)

Communication was established successfully on the first attempt. The CrashScan application identified the module,

confirmed the part number and software version, and completed the full EDR data download without errors. The resulting report was generated and documents two recorded crash events stored in the ACM at the time of the data extraction.

The successful communication confirms that the binary data transfer to the donor EEPROM was executed correctly — the module's firmware accepted and correctly parsed the recovered crash data, validating both the integrity of the SEM extraction and the correctness of the data transfer methodology.

### B. Module and Event overview

The CrashScan EDR report documents data from ACM. The module download was performed at ignition cycle 50,449. Two crash events are recorded and presented in Table 1.

TABLE I. CRASH EVENTS

| ***Parameter*** | ***Most Recent Event (Accident)*** | ***1st Prior Event (Historical)*** |
|---|---|---|
| Event Number | 2 | 1 |
| Event Type | Frontal | Frontal |
| Ignition Cycle at Event | 26,577 | 21,877 |
| Cycles Before Download | 23,872 cycles ago | 28,572 cycles ago |
| Odometer at Event | 562,311 km | 451,211 km |
| Operating Time at Event | 1,994,697 min | 1,371,312 min |
| Airbag Warning Light | ON | OFF |
| Max Longitudinal Delta-V | −20.00 km/h | −2.00 km/h |
| Time at Max Long. Delta-V | 170.0 ms | 105.0 ms |
| Max Lateral Delta-V | 3.00 km/h | 1.00 km/h |
| Time from Pre-Crash to Trigger | 124 ms | 465 ms |

### C. Most Recent Event — Crash Pulse Analysis

The crash pulse for the forensically relevant Most Recent Event is characterised by a progressive longitudinal deceleration developing smoothly from 0 km/h at trigger time (t = 0 ms), reaching a plateau at approximately −20 km/h between 170 ms and 250 ms. The shape of this pulse — a gradual build-up without abrupt initial spike — is consistent with progressive structural engagement in a frontal or predominantly frontal collision, in which the vehicle's crumple zone absorbs energy over an extended deceleration phase. Table 2 shows CrashScan EDR report: Longitudinal crash pulse data — Most Recent Event. Maximum delta-V of −20 km/h reached at 170 ms and sustained through 250 ms. Lateral crash pulse data — Most Recent Event (selected intervals). Maximum lateral delta-V 3 km/h at 100–120 ms.

The lateral delta-V remains modest throughout, with a maximum of 3 km/h recorded at 100–120 ms, confirming a predominantly longitudinal (frontal) impact character with only a minor lateral component. The side algorithm started at 63 ms, while the frontal algorithm activated at time zero — consistent with initial frontal structural engagement followed by secondary loading.

TABLE II. CRASH PULSE DATA

| ***Time (ms)*** | ***Longitudinal Delta-V (km/h)*** | ***Time (ms)*** | ***Longitudinal Delta-V (km/h)*** | ***Time (ms)*** | ***Longitudinal Delta-V (km/h)*** |
|---|---|---|---|---|---|
| 0 | 0 | 90 | −13 | 180 | −20 |
| 10 | 0 | 100 | −15 | 190 | −20 |
| 20 | −1 | 110 | −16 | 200 | −20 |
| 30 | −2 | 120 | −18 | 210 | −20 |
| 40 | −3 | 130 | −18 | 220 | −19 |
| 50 | −4 | 140 | −19 | 230 | −20 |
| 60 | −6 | 150 | −19 | 240 | −20 |
| 70 | −8 | 160 | −19 | 250 | −20 |
| 80 | −11 | 170 | −20 | | |
| ***Time (ms)*** | ***Lateral Delta-V (km/h)*** | ***Time (ms)*** | ***Lateral Delta-V (km/h)*** | ***Time (ms)*** | ***Lateral Delta-V (km/h)*** |
| 0 | 0 | 90 | 2 | 180 | 1 |
| 10 | 0 | 100 | 3 | 190 | 1 |
| 20 | 0 | 110 | 3 | 200 | 1 |
| 30 | 1 | 120 | 3 | 210 | 1 |
| 40 | 1 | 130 | 2 | 220 | 1 |
| 50 | 1 | 140 | 2 | 230 | 1 |
| 60 | 1 | 150 | 1 | 240 | 1 |
| 70 | 1 | 160 | 1 | 250 | 1 |
| 80 | 2 | 170 | 1 | | |

### D. Seat Belt and Airbag Status — Most Recent Event

The occupant restraint status recorded at −1.0 second prior to trigger shows the Passenger Seat Belt Status as Buckled, while the Driver Seat Belt Status is recorded as Unavailable — a data field absence that may reflect a damaged sensor circuit in the module. No airbag deployments were recorded for either driver or passenger positions across all airbag stages (front, torso, curtain). Neither driver nor passenger pretensioners deployed.

The Airbag Warning Light status recorded at −1.0 second was ON, indicating that the SRS system had a pre-existing fault condition registered prior to the collision event. This is forensically significant: it confirms that the ACM's warning lamp circuit was functional at the time of the crash trigger, and that the fault may have predated the traffic accident.

The investigation established that the seat belt pretensioners from a previous collision had not been replaced during subsequent repairs. This finding was confirmed by the workshop diagnostic history and by the identification labels on the seat belt assemblies. Consequently, the SRS control module contained a pre-existing fault, as corroborated by the Airbag Warning Light being ON prior to the collision. The available evidence indicates that this fault condition was the reason why the driver's airbag was not deployed during the investigated collision.

### E. Pre-Crash Data Analysis — Most Recent Event

The pre-crash data records vehicle dynamics over the 5.0 seconds immediately preceding the collision trigger, sampled at 0.5-second intervals. This dataset provides the most forensically compelling evidence in the entire EDR report.

Table 3 shows CrashScan EDR report: complete pre-crash data — Most Recent Event (forensically relevant accident).

TABLE III. PRE-CRASH DATA

| *Time (sec)* | *Vehicle Speed (km/h)* | *Engine Speed (RPM)* | *Accele-rator Pedal (%)* | *Brake Status* | *Steering Angle (deg)* | *Anti-Lock Brake System* | *Stability Control* |
|---|---|---|---|---|---|---|---|
| −5.00 | 54 | 1600 | 0.0 | Off | −62.0 (Left) | Off | On |
| −4.50 | 53 | 1536 | 0.0 | On | −56.0 (Left) | Off | On |
| −4.00 | 48 | 1664 | 0.0 | On | −24.0 (Left) | Off | On |
| −3.50 | 43 | 1536 | 0.0 | On | −8.0 (Straight) | Off | On |
| −3.00 | 37 | 1344 | 0.0 | On | −24.0 (Left) | Off | On |
| −2.50 | 33 | 1088 | 0.0 | On | −2.0 (Straight) | Engaged | On |
| −2.00 | 29 | 960 | 0.0 | On | 14.0 (Right) | Engaged | On |
| −1.50 | 28 | 832 | 0.0 | On | 92.0 (Right) | Engaged | On |
| −1.00 | 25 | 768 | 0.0 | On | 118.0 (Right) | Engaged | On |
| −0.50 | 22 | 640 | 0.0 | On | 18.0 (Right) | Engaged | On |
| 0.00 | 19 | 320 | 0.0 | On | 26.0 (Right) | Engaged | Engaged |

The pre-crash data documents a continuous driver response during the five seconds preceding the collision. The vehicle was initially travelling at 54 km/h and underwent progressive deceleration throughout the recorded interval. Braking commenced approximately 4.5 seconds before the impact and remained continuously applied until the collision. The recorded speed decreased to 19 km/h at the trigger point, corresponding to a total speed reduction of 35 km/h. The ABS system became active during braking, indicating that the available tire-road adhesion limit had been reached. Significant steering inputs were recorded in the final two seconds before impact, consistent with an evasive or corrective steering manoeuvre. At the moment of collision, both ABS and ESC were actively engaged, confirming that the vehicle stability systems were intervening. The recorded data indicate that the driver combined intensive braking with steering corrections in an attempt to avoid the collision. The collision occurred after a substantial reduction in vehicle speed. Overall, the pre-crash data demonstrate a sustained and coordinated driver response consistent with the perception of an imminent collision hazard.

### *F. 1st Prior Event — Historical Record*

The 1st Prior Event (ignition cycle 21,877, odometer 451,211 km) is a low-severity frontal event with a peak longitudinal delta-V of −2 km/h — well below the airbag deployment threshold and forensically consistent with a minor impact or aggressive braking event. The airbag warning lamp was OFF at the time of this event, and both driver and passenger pretensioners deployed at 4 ms — an indication that the event exceeded the pretensioner threshold but not the airbag threshold. The pre-crash data for this event shows the vehicle accelerating from 38 km/h to 45 km/h under full throttle (100%), followed by rapid deceleration with a steering input reaching −254° (extreme left). ABS was engaged for the final 2 seconds. This historical event is noted for completeness but is not the subject of the current forensic investigation.

## VII. DISCUSSION

The successful recovery of forensically usable EDR crash data from a physically destroyed, non-communicative ACM via SEM charge imaging represents a significant advance in the capabilities of forensic accident reconstruction. To the authors knowledge, this is the first documented case in which this complete workflow — from SEM imaging of a damaged automotive EEPROM through binary data transfer to a donor unit and final EDR report extraction using a standard forensic tool — has been executed successfully in a real criminal investigation context.

The forensic implications are substantial. Until now, a non-communicative ACM represented an absolute barrier to EDR data recovery: the investigator could document the module's failure but could not retrieve its contents. This case demonstrates that this barrier is not absolute — it is instead a function of the state of analytical technology and the availability of specialist expertise. Modules previously considered to contain irrecoverable data may be candidates for SEM-based recovery, provided their Flash EEPROM memory arrays have not been directly damaged.

The laboratory infrastructure required for this work — high-resolution SEM, precision polishing equipment, NIR microscopy, and semiconductor failure analysis expertise — represents an investment exceeding several hundred thousand euros and demands a level of specialist training that falls well outside standard forensic laboratory provision. This does not diminish the significance of the result; it defines the operational context in which this methodology is applicable and the type of case justifying its deployment.

In the present investigation, Collision Sciences CrashScan served as an independent forensic decoding platform, converting the recovered EEPROM contents into a standardised EDR report. The successful generation of a complete CrashScan report independently confirmed that the recovered binary data were authentic, internally consistent and fully compatible with the original ACM, thereby providing an additional level of technical validation beyond the manual interpretation of the hexadecimal data.

The consistency and reliability of the data acquired using CrashScan have been repeatedly validated in previous studies involving Mercedes-Benz vehicles through comprehensive comparative analyses against data retrieved with the Xentry diagnostic platform. These validation studies consistently demonstrated an excellent correlation between the datasets obtained by both approaches, confirming the accuracy and robustness of the CrashScan methodology. The present investigation further substantiates these findings, as the recovered CrashScan data exhibited complete consistency with the corresponding Xentry diagnostic records. This agreement confirms the high precision, reliability, and forensic validity of CrashScan for EDR data recovery and interpretation, even in demanding forensic scenarios.

Although Collision Sciences CrashScan was ultimately used to generate the final EDR report, the recovered EEPROM binary can also be analysed directly without the use of proprietary decoding software. Through many years of forensic examination of Mercedes-Benz Airbag Control Modules and detailed reverse engineering of their EEPROM memory organisation, the author has developed the capability to identify and interpret the principal EDR parameters directly from the raw hexadecimal memory image.

The methodology used by the author for direct interpretation and decoding of raw EEPROM hexadecimal data is based on many years of forensic research, reverse engineering, and practical analysis of Mercedes-Benz ACM memory structures. The detailed know-how underlying this decoding process, including the identification of memory layouts, proprietary data structures, parameter locations, and reconstruction algorithms, is not disclosed in this publication. Only the validated forensic results relevant to the present investigation are presented, while the underlying decoding methodology remains proprietary to the author.

## VIII. CONCLUSION

This paper describes fully working and successful method of extracting crash data from severely damaged SRS module. As a result of an accident its internal memory chip was destroyed beyond possibility to read out its data. This is the first documented case of successful extraction from severely damaged Flash EEPROM memory device with 100% success.

This investigation demonstrates that a non-communicative ACM should no longer be regarded as containing irrecoverable EDR evidence. SEM-based recovery enables reconstruction of the original EEPROM contents, while direct interpretation of the recovered hexadecimal data provides an independent forensic assessment of the recorded event. Collision Sciences CrashScan subsequently verified the recovered EEPROM by successfully decoding the reconstructed memory and generating a complete, standardised EDR report. The successful operation of CrashScan after SEM-based recovery confirms that physically destroyed ACMs can still yield complete and legally defensible EDR evidence. The combination of SEM recovery, expert interpretation of raw hexadecimal EEPROM data, and independent verification using Collision Sciences CrashScan substantially extends the practical capabilities of forensic accident reconstruction and establishes a robust methodology for future investigations involving non-communicative Airbag Control Modules.

In addition to software-based decoding, the recovered hexadecimal EEPROM image can be analysed directly. Through many years of forensic examination and reverse engineering of Mercedes-Benz ACM memory structures, the author has developed the capability to identify the principal EDR parameters directly from the raw hexadecimal data. Although Mercedes-Benz distributes EDR-related information across numerous non-contiguous memory locations to increase the complexity of data interpretation, the underlying structure can be reconstructed through detailed knowledge of the memory architecture. Consequently, essential crash parameters — including vehicle speed, braking status, steering angle, delta-V values, restraint system status and other key EDR variables — can be extracted directly from the EEPROM image without the use of proprietary decoding software. This capability enables forensic evaluation even before the data are processed by external software. In the present case, the manually interpreted hexadecimal data corresponded with the EDR report independently generated by Collision Sciences CrashScan, providing an additional level of technical verification.

Although the data stored inside the EEPROM array were not encrypted at the hardware level, neither scrambling nor encryption would stop the recovery. For example, some flaws inside embedded memory controllers could lead to data leakages [33], or microprobing could be used to defeat the encryption [24].

The presence of error correction helps not only to increase the success rate but also proves that no data were lost because of damage to the silicon chip. Most automotive chips have embedded memory with error correction to increase reliability.

The SPI EEPROM chip described in this work was fabricated with 130 nm process. Smaller fabrication process could pose more challenges and would require more sophisticated sample preparation and imaging process. Also, relatively small memory array of 16 kilobytes made the extraction process faster. For large memory arrays not only more advanced imaging techniques would be required, but also some form of automation to speed up the process. However, as long as all memory cells inside the array are not physically damaged there is a good chance to recover all the data without any errors.

This paper documents, for the first time in the forensic literature, the complete end-to-end recovery of automotive Event Data Recorder (EDR) crash data from a physically destroyed, non-communicative Airbag Control Module (ACM) using Scanning Electron Microscope (SEM) charge imaging of the EEPROM silicon die as the primary data extraction technique. This could find applications not only in automotive but also in medical, aviation and aerospace industries if semiconductor devices were mechanically, electrically or fire damaged.